\documentclass[
aps,
prab,
reprint,
twocolumn,
superscriptaddress]
{revtex4-2}

\usepackage{amssymb}
\usepackage{amsthm}
\usepackage{fancyhdr}
\usepackage{graphicx}
\usepackage[colorlinks, linkcolor=green!40!black, citecolor=blue!80!black]{hyperref}
\usepackage[table]{xcolor}
\usepackage{multirow}
\usepackage{rotating}
\usepackage{float}
\usepackage{array}
\usepackage{supertabular}
\usepackage{amsmath}
\usepackage{dcolumn}
\usepackage{tipa}
\usepackage{siunitx}
\usepackage{bm}
\usepackage{booktabs}
\usepackage[capitalise]{cleveref}
\usepackage{optidef}
\usepackage{tabularx}

\newcommand{\optf}[1]{{\cellcolor[gray]{0.8}} #1}

\providecommand{\abs}[1]{\lvert#1\rvert}
\providecommand{\norm}[1]{\lVert#1\rVert}

\begin{document}

\title{Data-Driven Gradient Optimization for Field Emission Management\texorpdfstring{\\}{} in a Superconducting Radio-Frequency Linac}

\author{S. Goldenberg}
\email{sgolden@jlab.org}
\affiliation{Thomas Jefferson National Accelerator Facility, Newport News, Virginia 23606, USA}
\author{K. Ahammed}
\affiliation{Old Dominion University, Norfolk, Virginia 23529, USA}

\author{A. Carpenter}
\affiliation{Thomas Jefferson National Accelerator Facility, Newport News, Virginia 23606, USA}
\author{J. Li}
\affiliation{Old Dominion University, Norfolk, Virginia 23529, USA}
\author{R. Suleiman}
\author{C. Tennant}
\affiliation{Thomas Jefferson National Accelerator Facility, Newport News, Virginia 23606, USA}

\date{\today} 

\begin{abstract} 
    Field emission can cause significant problems in superconducting radio-frequency linear accelerators (linacs).
    When cavity gradients are pushed higher, radiation levels within the linacs may rise exponentially, causing degradation of many nearby systems.
    This research aims to utilize machine learning with uncertainty quantification to predict radiation levels at multiple locations throughout the linacs and ultimately optimize cavity gradients to reduce field emission induced radiation while maintaining the total linac energy gain necessary for the experimental physics program.
    The optimized solutions show over 40\% reductions for both neutron and gamma radiation from the standard operational settings.
\end{abstract}

\maketitle

\section{Introduction}

Jefferson Lab's Continuous Electron Beam Accelerator Facility (CEBAF)~\cite{PhysRevAccelBeams.27.084802} relies on two superconducting radio-frequency linear accelerators (SRF linacs) to deliver high-energy electron beams to nuclear physics experiments in the four experimental halls~\cite{ARRINGTON2022103985}.
An integral part of these linacs are cryomodules which contain multiple SRF cavities.
These SRF cavities provide the main accelerating gradients to the electron beam, and currently produce the \SI{12}{GeV} beam necessary for scientific discovery.

\subsection{Field Emission}

When SRF cavities are operated at high
radio-frequency (RF) gradients, it is possible for electrons to be emitted from the cavity walls,
known as field emission 
(FE)~\cite{Fowler:1928bv,Padamsee:2008}. 
The basic physics of FE in SRF cavities is generally understood \cite{10.1063/1.1663338,PhysRevSTAB.16.112001,Reschke:2017gjp}, and modern cavity surface 
processing and assembly techniques control FE well prior to installation~\cite{Omet:2023dfd}.
However, during routine operation, degradation of the FE onset ({\it i.e.}, the gradient at which the first FE-induced gammas are detected) has been observed.
It is believed this degradation is caused by particulates and hydrocarbons entering the cavity, contaminating its surface, 
and introducing new field emitters~\cite{martinello2022plasmaprocessinginsitufield}.

Field emission is one of the most detrimental problems for CEBAF linacs \cite{Drury:2013kva, Freyberger:IPAC2015-MOXGB2,Geng:IPAC2016-THOBB03,Geng:IBIC2017-TH1AB1}.
When field emitted electrons hit the cavity walls, cryogenic heat loads increase.
If those electrons are accelerated beyond the source cavity, their energies can exceed neutron production and material activation thresholds, producing radiation and damaging beamline components.
Prior to the~\SI{12}{GeV} upgrade to CEBAF, activation levels detected were typically low enough to not require ``Radiation Area'' postings.
These postings are required for areas where the whole-body radiation dose is $>$\SI{0.005}{rem\per\hour}.
However, since the energy upgrade, CEBAF has suffered from significant FE induced radiation.
With RF on, dose rates observed 
at~\SI{30}{\centi\meter} from the beamline are as high as~\SI{10}{rem\per\hour} and~\SI{100}{rem\per\hour} for neutron and gamma radiation, respectively.
This level of radiation causes significant damage to beamline components, including vacuum valves, magnets, and cables of beam position monitors and ion pumps.  
Replacing these components can use significant resources.
Worse, portions of both linacs are considered ``Radiation Areas'' for days or even weeks into scheduled downtime, limiting maintenance activities to repair and replace components, and increasing radiation doses for personnel.

Most vacuum valves in both linacs are now damaged causing them to leak. Radiation damaged valves, where the valves' seals become brittle might be one of the 
most detrimental problems preventing higher cavity gradients. Particulates from the damaged seals propagate through the beamline, entering the cryomodules. 
This contamination introduces FE even at very low gradients and requiring the removal and the refurbishments of the cryomodules. 
The leaky valves prevent vacuum isolation of the cryomodules, complicating the maintenance activities and the removal and installation of cryomodules.
Pending available funds, CEBAF has started on a multi-year project to replace Viton vacuum valves in the linacs with all-metal valves~\cite{VAT.vacuum-valves}. 
The refurbishment is further complicated by the activation of many of the cryomodule and beamline components due to FE.

Determining FE-source cavities is challenging for a number of reasons. At CEBAF, existing radiation monitors were 
designed to detect high bursts of gamma radiation due to beam loss events and quickly shutdown the beam.
Until recently, there was no readily available way to continuously monitor the relatively low level of radiation produced by FE in real-time within the accelerator. 
Newly designed neutron and gamma radiation dose rate detectors (NDX)~\cite{ndxmeters,degtiarenko2019neutron} were successfully installed and provide real-time radiation measurements in 13 locations along both linacs at CEBAF.
However, detecting FE at the cavity level remains a challenge as each set of 13 detectors covers a total of 200 cavities.
Additionally, field emitted electrons may be captured and accelerated from the source cavity, causing radiation far away from the source. 
Tests at CEBAF have shown that FE can be accelerated in either direction in the linac, with one instance passing through 14 cryomodules ($>$\SI{100}{m})~\cite{Geng:IBIC2017-TH1AB1}.

\begin{figure}
    \centering
    \includegraphics[trim={7.5cm 2cm 4cm 3.5cm},clip,width=1.0\linewidth]{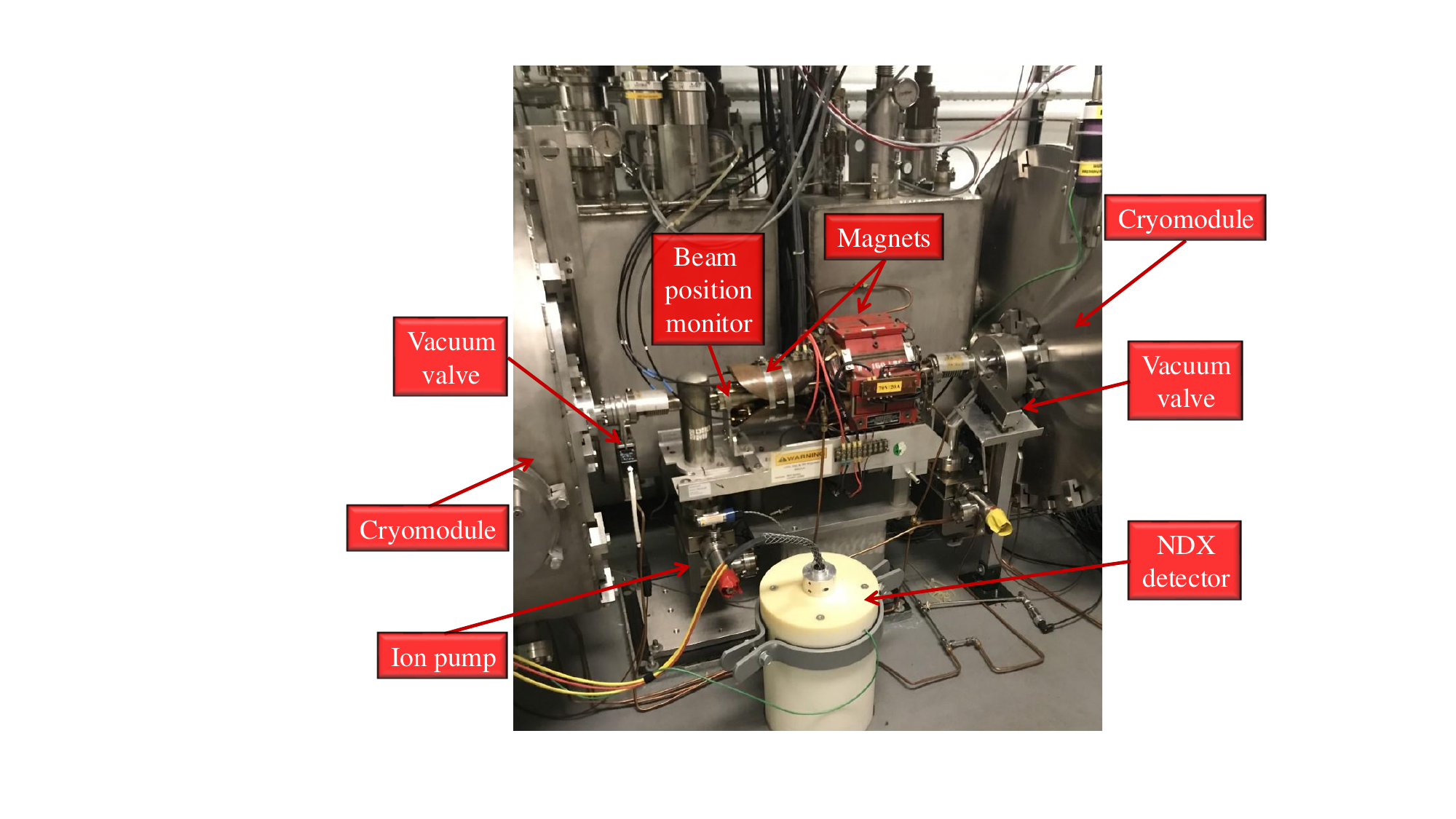}
    \caption{NDX 2L25 detector positioned next to the beamline connecting two cryomodules at a distance of about \SI{0.5}{m}. Neighbouring cryomodules are separated by a \SI{1.2}{m} long beamline. The white polyethylene outer layer of the detector is yellowing due to FE radiation.}
    \label{fig:NDX-detector}
\end{figure}

Figure~\ref{fig:NDX-detector} shows one of the NDX detectors positioned between two cryomodules at a distance of about \SI{0.5}{m} from the beamline. 
At the center of the detector, two identical ionization chambers~\cite{LND-ionization-chambers} filled with different helium gas isotopes ($^{3}\text{He}$ and $^{4}\text{He}$) serve as the sensors in the detector. 
The chambers' output currents are measured by highly sensitive electrometers~\cite{I400-electrometer}.  This configuration enables accurate measurement of the two different radiation dose rates.

\begin{figure}
    \centering
    \includegraphics[width=0.95\linewidth]{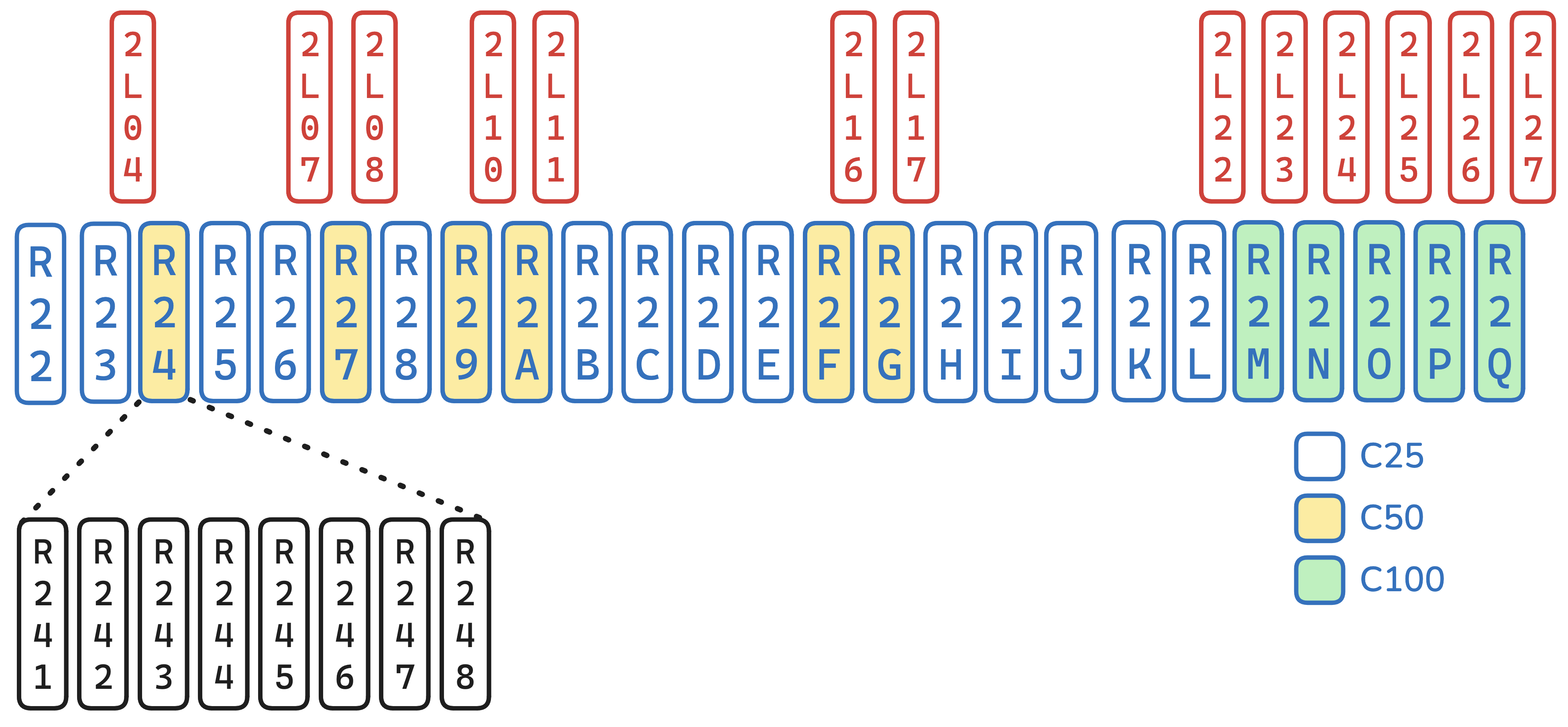}
    \caption{Layout of the CEBAF south linac. Cryomodules are outlined in blue, detectors in red. There are 8 individual cavities per cryomodule (shown in black). The south linac contains three cyromodule types indicated by their background color: C25, C50, and C100, where the numbers indicate the nominal energy gain in \SI{}{MeV} expected from each type.}
    \label{fig:detector_locations}
\end{figure}

A diagram of the south linac including the locations of the 25 cryomodules (containing eight cavities each) and 13 detectors is given in \cref{fig:detector_locations}.
These detector locations correspond to cryomodules with expected FE activity.  Cryomodule styles are typically denoted based on the expected total energy gain.  A C25 cryomodule, for example, would be expected to produce around 25 MeV of energy gain.
In CEBAF's south linac, detectors have been placed near the C50 and C100 cryomodules which produce its highest gradients and are therefore more likely to produce FE.

\subsection{Methodology}

If an FE source cavity can be determined, radiation levels can be minimized by lowering its gradient and redistributing gradient to other FE-free cavities. 
A proof-of-principle test of this gradient redistribution for reducing FE radiation was demonstrated in the injector by slowly increasing the gradient for each cavity individually to determine its contribution to the radiation levels~\cite{Geng:IBIC2017-TH1AB1}.
Individual testing like this is not feasible as a long-term solution, as it requires a significant investment in time. 
Additionally, it is very likely for field emitters to be enhanced and new field emitters to be excited, causing previous FE onset values to change.

These changing onset values, as well as changes in field emission intensity, make the task of modeling radiation responses particularly challenging. 
Furthermore, a cavity's contribution to radiation levels is rarely isolated, as interactions with nearby cavity gradients can alter radiation response by accelerating FE upstream or downstream.
These complex interactions and the high dimensional input space of the problem make it an ideal problem for machine learning (ML). 
Our goal is to produce an ML surrogate model that predicts radiation levels based on cavity gradients. 
This model will power an offline optimization process for redistributing gradients to reduce radiation levels while maintaining linac energy gain. This will be the first time ML is used to mitigate the problem of field emission in an SRF linac with a great number of cavities. 
Other ML efforts at CEBAF to improve SRF operation are described in~\cite{Rahman_2024, ferguson:ipac2024-tups69, PhysRevAccelBeams.23.114601}, while work at other SRF facilities can be found in~\cite{PhysRevAccelBeams.26.012801,boukela2024twostagemachinelearningaidedapproach,wang2023enhancing}.

\section{\label{sec:data}Data}

Our models utilize two sources of data, the RF control system and the NDX system. 
From the RF system we use the measured cavity gradient (MV/m) for model input. Other miscellaneous values that describe the broader machine state are also used to validate our data collection. The cavity gradients are near instantaneous measurements and updated at a rate of 1 Hz.  While CEBAF has multiple styles of cryomodules and RF controls, the data we use is common across all SRF cavities within a linac.

NDX detectors measure both neutron and gamma radiation dose rates over a specified time horizon. 
We choose to measure dose rates over one second windows as it matches the RF gradient update rate, provides a reasonable trade-off between response time and measurement variability, and to match the standard configuration used during beam operations. 
This system provides the 26 radiation dose rates (neutron and gamma for each of the 13 detectors) used as targets when training our models.

Since our goal is to minimize radiation during experimental operations, we require that our data be as reflective of those machine states as possible.  
This means we want all RF cavities to be properly phased, to have gradient settings similar to their operational gradients, and with beam transport magnets set to match operational optics for those gradients.
However, this does not require that all cavities be operational and running at their design limits. 
With 200 SRF cavities per linac, beam operations routinely occur with some degraded or disabled cavities. 
Additionally, some cavities are intentionally run below their design limits in order to reduce the frequency of certain RF trip types.  
This however gives us headroom for rebalancing gradient throughout a linac both during data collection and beam operations.

\subsection{\label{subsec:data_collection}Data Collection}

For this work, we developed software that actively controls a linac to investigate the radiation response to changing SRF cavity gradients.
Raw data was collected by first configuring a linac to operational standards with RF on and beam absent. 
Collecting data without the beam should not bias our results, as under normal operation, the electron beam does not influence radiation levels in the linac unless there is a beam loss event. 
Beam loss events do not produce the constant radiation responses that require mitigation through gradient optimization as they result in a large spike in radiation levels for a very short time, generally less than \SI{50}{\micro\second}, and cause the beam to shut down.

Our software scans a variety of gradient settings for individual cavities or for combinations of cavities in the neighborhood of their typical set points.
Individual cavity scans produce results that are easy for human analysis and can clearly identify which cavities are active field emitters.
However, individual cavity scan data lacks information on radiation responses when many cavities are changed at once.
Additionally, individually scanning 200 separate cavities is a time intensive endeavor.
On the other hand, combination scans demonstrated two significant advantages in our testing. 
First, combination scan data more closely approximates solutions produced by our optimizers as they alter multiple cavity gradients to reduce radiation and rebalance energy gain throughout the linac. 
Second, due to sparse opportunities for data collection, it was important to efficiently sample a wide variety of gradient distributions.
Adequate coverage of simultaneous gradient changes is critical for training models that perform well in the optimization phase.

Data collection is challenging in practice due to two main factors. 
First, our sampling techniques often exercised the RF systems harder than steady state beam operations, and second, the 200 cavity gradients created an enormous sample space which required being strategic with our data collection to make the best use of the allocated time.
Ensuring sufficient coverage in this environment within our 0.5-4 hour window required both forethought and flexibility.

Most data collection used combination scans with gradient offsets ranging from 
approximately~\SIrange[retain-explicit-plus, per-mode=symbol]{-4}{+0.5}{\mega\volt\per\meter}.
The exact distributions varied, but roughly 20\% of cavities experienced large gradient reductions, and 10\% of cavities remained completely unchanged. 
Some scans only introduced changes to every fifth cryomodule to reduce correlations between neighboring cavities. 
Smaller individual cavity scans of suspected field emitters would be performed if time permitted.

Early data collection efforts made all gradient changes simultaneously and in parallel, which caused unwanted correlations in our training data.
This was improved for our final datasets by introducing a random delay before changing a cavity gradient.
These delays were sampled from a uniform $\mathcal{U}(0, x)$ distribution where $x$ ranged from 30-60 seconds depending on the scan. 
Changes to individual cavity gradients were effected slowly, between 0.1-0.4 MV/m per second, in order to maintain RF stability and track changes in radiation levels. 
Cavities would take additional pauses as needed to accommodate supporting systems such as RF tuners or the cryogenic system.

After data collection, the raw data was processed by removing known issues.
This included removing samples where cavities had experienced an RF fault.
Finally, NDX detectors provide average dose rates for the preceding one second window which lag gradient readings and require adjustment.

In total, we collected data on three occasions.
Information about these datasets is included in \Cref{tab:datasets}.
Some sessions were devoted to data collection, while others involved short tests of individual settings (labeled scan and demo respectively). 
Neutron radiation readings appear relatively stable over this time frame, however gamma radiation has a more noticeable increase in the final data sets.

\renewcommand{\arraystretch}{1.2}
\setlength{\tabcolsep}{0pt}
\begin{table}[]
    \centering \footnotesize
    \sisetup{round-mode=places, round-precision=2, table-alignment=right, table-format=2.2}
    \caption{Statistical and categorical information about our data. Neutron and Gamma radiation readings (in rem/h) were totaled over all 13 detectors and averaged over the first 10 samples of that day's data collection to get initial readings.}
    \begin{tabular*}{0.95\linewidth}{@{\extracolsep{\fill}} llS[table-format=5, round-mode=none] SS[table-format=3.2]} \toprule
        Date & Type & {Samples} & {Neutron} & {Gamma} \\ \midrule
        May 9    & Scan & 5880  & 17.3840 & 228.1175 \\ 
        May 14   & Scan & 4134  & 16.4435 & 219.6261 \\ 
        May 19 & Scan & 1733 & \multicolumn{1}{r}{\multirow{2}{*}{18.82}} & \multicolumn{1}{r}{\multirow{2}{*}{283.82}} \\
        May 19   & Demo  & 1200  &  &  \\ \bottomrule
    \end{tabular*}
    \label{tab:datasets}
\end{table}

\subsection{\label{subsec:drift}Data Drift}

\begin{figure}[t]
    \centering
    \includegraphics[width=1.0\linewidth]{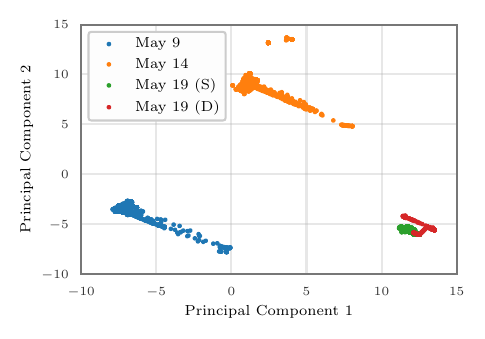}
    \caption{Example of data drift. PCA-based dimensionality reduction of measured cavity gradients from the different data sets we collected. (S) and (D) denote the scan and demonstration datasets, respectively.} 
    \label{fig:data_drift}
\end{figure}

The field emission and radiation environment in CEBAF is prone to both sudden and gradual change.
This is due to changes in the configuration of CEBAF, as well as changes in radiation response to a given configuration. 
During an experimental run, SRF cavities routinely experience problems that reduce their gradient or require the cavity to be completely bypassed.
In most cases, these problems are addressed on the next maintenance period. 
In the interim, the lost gradient must be redistributed throughout the linac to maintain the linac energy gain required by experimenters.  
Given the complexity of modern particle accelerators, a linac likely has many other unidentified sources of drift.
We use principal component analysis (PCA) to reduce the gradient sample space for the south linac's 200 cavities down to two dimensions.
These two dimensions clearly show the aggregate change of cavity gradients across the different datasets we use (see \cref{fig:data_drift}). 
The behavior of field emitters also changes routinely.  
We have observed both sudden turn-on events where a field emitter rapidly strengthens producing much higher radiation than before, and more gradual changes where field emitters slowly change their average radiation level over the course of hours or days without major changes to cavity gradients.

This paper primarily focuses on modeling of radiation over short time scales without the use of continual learning strategies.  However, even at the time scale of a week, we observe the radiation signature changing at individual detectors.

\section{\label{sec:models}Surrogate Modeling}

The ability to perform offline optimization necessitates the ability to predict radiation levels throughout a linac given a set of cavity gradients. 
Change is a constant in CEBAF's linacs; our solution must also allow for simple, periodic re-training. 
Methods that provide uncertainty quantification are desirable to help to guide an optimizer away from under performing regions of the model based on noisy or poorly sampled regions of data. 
There are a variety of methods that could accomplish these goals, including Quantile Regression methods \cite{koenker2005QRBook, shen2021deep, chronopoulos2023forecasting}, Bayesian methods and their approximations like MC-Dropout \cite{gal2016dropoutUQ}, Gaussian Process methods like SNGP \cite{liu2020SNGP}, and Ensemble methods like Deep Ensembles \cite{lakshminarayanan2017deepensembles}.
A more complete review of uncertainty quantified deep learning can be found in \cite{abdar2021UQreview}.

As our first attempt to improve FE-induced radiation, we focused our surrogate modeling efforts on the south linac using a Simultaneous Quantile Regression (SQR) model \cite{tagasovska2019SQR}. 
Our work focused on this approach due to its inference speed, support of uncertainty quantification, ease-of-update, and ability to support high dimensional inputs.

\subsection{\label{subsec:sqr}Architecture}

To reduce the possibility of data and model correlations affecting performance, we developed a single model per detector where the input for each model consists of 16 gradients corresponding to the cavities in the two adjacent (upstream and downstream) cryomodules.
While this could cause problems for instances where FE travels further from the source cavity, the majority of the radiation response is often seen locally and therefore will be captured by this reduced model.
We found this approach was more likely to match the FE behavior when compared to a single model using all 200 gradient measurements.

Each of our neural network models utilize SQR to produce predictions as well as an estimate of model uncertainty.
SQR is an extension of Quantile Regression which attempts to model functions that represent specific quantiles by minimizing the quantile loss.
Given a chosen quantile $0<\tau<1$, observations $y$, and predictions $\hat{y}$, the quantile loss over $n$ samples is defined by:  

\begin{equation}
    \frac{1}{n} \sum_{i=0}^{n}(y_i-{\hat{y}}_i)(\tau - 1_{y_i<{\hat{y}}_i}) 
\end{equation} 

where $1_{y_i<\hat{y}_i}$ is the indicator function for $y_i< \hat{y}_i$.
This loss is not symmetric ({\it i.e.} positive and negative losses are treated differently) in order to push the convergence of the model away from the median and towards the desired quantile.
When $\tau=0.5$, using the quantile loss is equivalent to half the mean absolute error (MAE): $\sum_{i=0}^n \norm{y_i - \hat{y}_i} / n$.

Compared to standard Quantile Regression, SQR estimates the entire conditional distribution through a single model rather than separate models or outputs for each desired quantile.
This can help reduce a problem commonly known as the quantile crossing problem, where $\tau_1 < \tau_2$ but $f_{\tau_1}(x) > f_{\tau_2}(x)$ where $f_\tau(x)$ is the model prediction for input $x$ and quantile $\tau$.
To do this, a single model is trained using the quantile loss by including the desired quantile as part of the input to the model.
Therefore, the model learns a single conditional function $f(x,\tau)$ instead of multiple separate ones which is believed to be more likely to produce a monotonic function with respect to $\tau$ \cite{tagasovska2019SQR}.

To be specific, when training the SQR model, $\tau$ is chosen for each sample in each batch of data as a uniform random variable between 0 and 1, and the loss is calculated using those values of $\tau$.
For this work, we use $\tau = 0.5$ ({\it i.e.} $f(x, 0.5)$) for predictions as it estimates the median radiation response. 
Uncertainty estimates were produced by choosing $\tau =$ 0.16 and 0.84, which represents $\pm1$ standard deviations of a standard Gaussian random variable.
Then we compute $\sigma = \abs{f(x, 0.84) - f(x, 0.16)} / 2$ for the estimated model uncertainty with gradient settings $x$.
By computing $\sigma$ this way, we provide a single value that represents the model uncertainty to pass to our optimizer and the Uncertainty Toolbox \cite{chung2021uncertainty} for calibration metrics.
One forward pass is required for each value of $\tau$, totaling 3 forward passes.
However, since $\tau$ is an input to the model, these forward passes can be batched together to improve prediction speed.

Our model architecture is shallow but wide, encompassing 16 inputs, a single dense hidden layer with 550 units, a Dropout layer with a rate of $0.4$, and a dense output layer with 2 units: one for neutron and the other for gamma radiation. 
Both dense layers were regularized using L2 regularization with a factor of \num{4e-5} and a \verb|leaky_relu| activation defined by 
\begin{equation}
    f(x) = \begin{cases}
    x,& \text{if } x\geq 0\\
    \alpha x,              & \text{otherwise}
    \end{cases}
\end{equation}
where $\alpha=0.3$. 
We explored using additional cryomodules but found no significant changes to model performance. 
These initial hyperparameters were chosen based on a less formal preliminary study in December 2023.
While it is possible our model could be improved with further hyperparameter tuning, the performance was sufficient to achieve good performance on our May training sets without over-fitting.

\subsection{\label{subsec:model_performance}Performance}
We preformed 8-fold cross validation for data collected on May 9.
First, the 5880 data points were split into 56 batches containing 105 seconds of data each.
These 56 batches were then randomly divided into 8 folds with 7 batches for testing.
Each fold is then trained on the remaining 49 batches for a total of 5145 training and 735 testing samples. 
This was done to ensure that each fold is tested with at least a few samples with novel gradient settings, as a purely sequential splitting created folds with either very low or very high variability in gradients.
We present the mean and standard deviation of the root mean squared error (RMSE) over the 8 folds for each sensor in \Cref{tab:kfold-results}.
Additionally, we show the standard deviation of each detector signal over the first 80 samples to show an estimate of the sensor noise when gradients are stable.

\renewcommand{\arraystretch}{1.2}
\setlength{\tabcolsep}{5pt}
\begin{table}[]
    \centering \footnotesize
    \sisetup{round-mode=places, round-precision=3, table-alignment=right, table-format=1.3}
    \caption{Mean and standard deviation of test RMSE (denoted $\mu$ and $\sigma$ respectively) over 8 folds. We also calculate the noise of each detector (denoted $\epsilon_n$ and $\epsilon_\gamma$) based on the first 80 samples of the May 9 data collection when gradients were stable. We see the models for all detectors before 2L22 have average RMSE values that match the data noise. The models for detectors near the C100s, which experience the largest changes in radiation, are less accurate.}
    \begin{tabular*}{0.95\linewidth}{@{\extracolsep{\fill}} lSSSSSS} \toprule
    Detector & {$\epsilon_n$} & {$\mu_n$} & {$\sigma_n$} & {$\epsilon_\gamma$} & {$\mu_\gamma$} & {$\sigma_\gamma$} \\ \midrule
    2L04     & 0.007885       & 0.008342  & 0.000267     & 0.464134            & 0.544971       & 0.035670          \\
    2L07     & 0.006940       & 0.007172  & 0.000219     & 0.496039            & 0.597381       & 0.049436          \\
    2L08     & 0.011387       & 0.013507  & 0.001954     & 0.601464            & 0.940764       & 0.211236          \\
    2L10     & 0.011879       & 0.014108  & 0.000224     & 0.486585            & 0.486782       & 0.013525          \\
    2L11     & 0.013548       & 0.014009  & 0.000318     & 1.923603            & 2.007512       & 0.035241          \\
    2L16     & 0.016464       & 0.016738  & 0.000483     & 0.870934            & 0.886105       & 0.029148          \\
    2L17     & 0.017175       & 0.016902  & 0.000701     & 2.241649            & 2.235415       & 0.098110          \\
    2L22     & 0.007530       & 0.052311  & 0.030051     & 0.501513            & 1.648745       & 0.622813          \\
    2L23     & 0.014355       & 0.276094  & 0.138392     & 1.313257            & 4.668636       & 1.940858          \\
    2L24     & 0.012697       & 0.258971  & 0.090777     & 0.507190            & 3.123450       & 0.932382          \\
    2L25     & 0.033036       & 0.310259  & 0.131493     & 0.506150            & 1.920767       & 0.813834          \\
    2L26     & 0.017987       & 0.374536  & 0.103127     & 0.808025            & 4.856736       & 1.493690          \\
    2L27     & 0.010008       & 0.100194  & 0.038288     & 0.637973            & 1.181415       & 0.258929          \\ \bottomrule
    \end{tabular*}
    \label{tab:kfold-results}
\end{table}

\Cref{tab:kfold-results} shows that our models are accurate for detectors near C25's and C50's (2L04-2L17) and do not overfit to the detector noise ($\epsilon$), while model predictions for detectors near the C100's are less accurate. 
This makes sense as little-to-no radiation was detected in the 2L04-2L17 regions.
The average RMSE for detectors which experience meaningful radiation is within 0.4 \unit{rem \per \hour} for neutron radiation, and 5 \unit{rem \per\hour} for gamma radiation.
Additionally, the standard deviation of RMSE over the 8 folds is much higher for these detectors, indicating that the test sets for certain folds may contain novel gradient responses that are hard to predict.

\begin{figure*}[ht]
    \centering
    \includegraphics[]{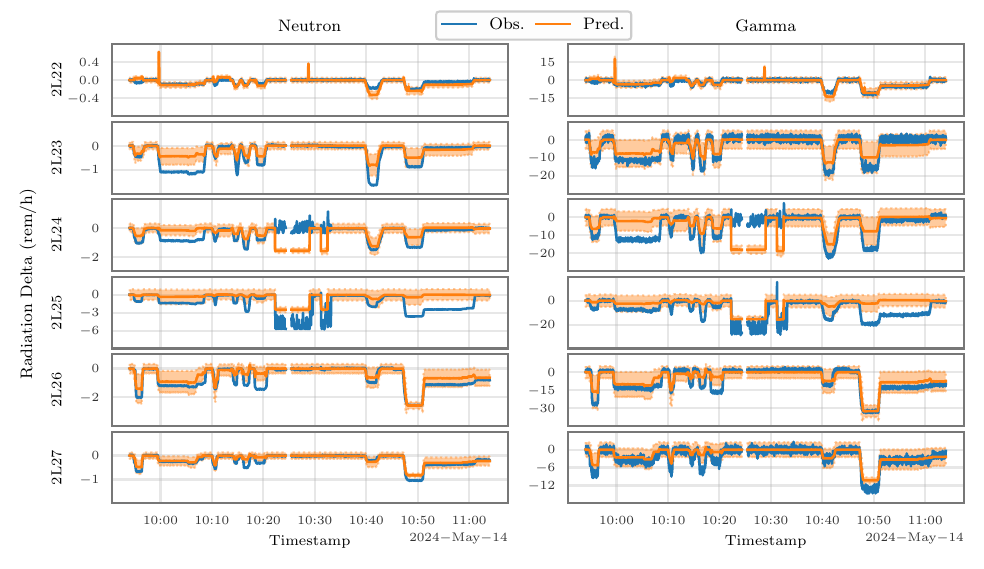}
    \caption{Radiation delta from the initial timestep for observations (in blue) and model responses (in orange) on  the May 14 dataset. Uncertainties are shown as an orange shaded region equivalent to $\pm 3 \sigma$. By plotting the radiation delta rather than the exact observations and predictions, we remove any systematic shift and highlight true missed predictions. While most model responses are well correlated, we see a possible change of behavior for 2L24 (near 10:25).}
    \label{fig:may14_results}
\end{figure*}

To better understand these results, we calculated the standard deviation over the full May 9 collection for each detector.
We found the standard deviation for 2L04-2L17 neutron radiation over all of May 9 is less then 0.017 \unit{rem \per \hour}, which matches the maximum $\epsilon$ for those detectors in \Cref{tab:kfold-results}.
This indicates that these detectors are dominated by sensor noise rather than variation as a result of gradient changes. 
However, the standard deviation for detectors 2L22-2L27 is more than an order of magnitude higher than the calculated noise floor (maximum 0.613). 
A similar phenomenon is observed for gamma radiation.
Therefore, we truncate our tables to focus on detectors 2L22-2L27 which show non-trivial gradient responses. 
For completeness, the tables for detectors 2L04-2L17 are presented in \Cref{appendix:2L04/17_Results}.

Before running a full test on the accelerator, we were able to collect data on May 14 to validate our model trained on all of May 9's data. 
RMSE results and Pearson correlation coefficients for this model and dataset are presented in \Cref{tab:c100-may14-results}.
We see large increases in average error for all but 2L23, however, all detectors except 2L24 have good correlation values.
To emphasize this, we plot \Cref{fig:may14_results}, which shows the radiation delta from the beginning of data collection on May 14. 
This plot removes any systematic shift in the predictions that could be accounted for prior to data collection, and helps to visualize the correlation between the observations and predictions.
We also show an uncertainty band covering $\pm 3 \sigma$ and see that most observations are contained within the band.
One notable portion of these graphs where the observations fall outside of the uncertainty band is between 10:20 and 10:35 where there was a large predicted reduction in radiation that did not occur in 2L24. 
However, that drop in radiation exists in 2L25 and is slightly under-predicted for neutron radiation, suggesting that the behavior of a field emitter may have become more directional or moved away from 2L24, pushing radiation further down the linac.

\renewcommand{\arraystretch}{1.2}
\setlength{\tabcolsep}{5pt}
\begin{table}[]
    \sisetup{round-mode=places, round-precision=3, table-alignment=center, table-format=1.3, table-space-text-pre = $-$}
    \centering \footnotesize
    \caption{May 14 RMSE and correlation for the neutron (n) and gamma ($\gamma$) predictions from our model trained on the full May 9 data. Other than 2L23, all detectors show a noticeable increase in RMSE when compared to \Cref{tab:kfold-results}. }
    \begin{tabular*}{0.95\linewidth}{@{\extracolsep{\fill}} lSSSS} \toprule
    \multirow{2}{*}{Detector} & \multicolumn{2}{c}{RMSE} & \multicolumn{2}{c}{Corr} \\
     & {n}      & {$\gamma$} & {n}      & {$\gamma$} \\ \midrule
    2L22     & 0.168764 & 2.566963   & 0.850233 & 0.913355   \\
    2L23     & 0.267386 & 2.861557   & 0.938215 & 0.945545   \\
    2L24     & 0.573765 & 6.726722   & 0.303461 & 0.395763   \\
    2L25     & 1.210199 & 6.570722   & 0.766477 & 0.725922   \\
    2L26     & 1.518077 & 18.06952   & 0.925594 & 0.924085   \\
    2L27     & 0.245420 & 2.224972   & 0.962038 & 0.928126   \\ \bottomrule
    \end{tabular*}
    \label{tab:c100-may14-results}
\end{table}

\subsection{\label{subsec:uq} Uncertainty Quantification}

To determine the calibration of our models, we utilize the expected calibration error (ECE) calculated using the Uncertainty Toolbox~\cite{chung2021uncertainty}.
This error is calculated by comparing normalized residuals, $(y_i-\hat{y_i})/\sigma_i$, with the inverse cumulative distribution function (CDF) of a standard Normal $\mathcal{N}(0,1)$ distribution.
Specifically, ECE calculates the average absolute error between the observed proportion of normalized residuals and the expected proportion predicted by a normal distribution for 100 different equally spaced proportions between 0 and 1. 

\renewcommand{\arraystretch}{1.2}
\setlength{\tabcolsep}{5pt}
\begin{table}[]
    \sisetup{round-mode=places, round-precision=3, table-alignment=center, table-format=1.3, table-space-text-pre = $-$}
    \centering \footnotesize
    \caption{ECE values using our model trained on the full May 9 data. We provide results for both the May 9 training data and the May 14 validation data. We see the model is generally well-calibrated in training, but performs worse in validation for the C100 detectors.}
    \begin{tabular*}{0.95\linewidth}{@{\extracolsep{\fill}} lSSSS} \toprule
    \multirow{2}{*}{Detector} & \multicolumn{2}{c}{Training} & \multicolumn{2}{c}{Validation} \\    
             & {n}      & {$\gamma$} & {n}      & {$\gamma$} \\ \midrule
    2L04     & 0.026063 & 0.058519   & 0.025140 & 0.035741   \\ 
    2L07     & 0.033924 & 0.060626   & 0.015868 & 0.054695   \\ 
    2L08     & 0.032117 & 0.057193   & 0.038995 & 0.052412   \\ 
    2L10     & 0.012715 & 0.084101   & 0.011200 & 0.074203   \\ 
    2L11     & 0.032471 & 0.033306   & 0.042638 & 0.033181   \\ 
    2L16     & 0.024162 & 0.031441   & 0.015385 & 0.019287   \\ 
    2L17     & 0.041050 & 0.029363   & 0.034534 & 0.024768   \\ 
    2L22     & 0.065799 & 0.027893   & 0.477747 & 0.407123   \\ 
    2L23     & 0.118568 & 0.027965   & 0.441976 & 0.174680   \\ 
    2L24     & 0.129277 & 0.066729   & 0.325062 & 0.334158   \\ 
    2L25     & 0.139834 & 0.061806   & 0.333258 & 0.390841   \\ 
    2L26     & 0.142169 & 0.101468   & 0.49     & 0.489995   \\ 
    2L27     & 0.153034 & 0.059095   & 0.486015 & 0.372319   \\ \bottomrule
    \end{tabular*}
    \label{tab:uq-may14-results}
\end{table}

The results of the May 9 model on the training data as well as the May 14 validation set are shown in \Cref{tab:uq-may14-results}.
It is clear that the model is relatively well-calibrated on the training data, but its performance deteriorates for the C100 detectors which aligns with the increases in RMSE in \Cref{tab:c100-may14-results}.
After further analysis, we noticed the standard deviation of our uncertainty estimates remained very similar for both datasets and the mean uncertainty estimates only increased by a small margin.
This leads us to believe that the model has done a good job modeling the detector noise, however, when predictions are less accurate (like those on the May 14 data), the model becomes less well-calibrated.
In general, these results reinforce the idea that providing accurate uncertainties for out-of-distribution data is especially challenging.
Additionally, if the relationship between radiation and gradients changes, i.e., concept drift exists, it may be nearly impossible for a model that only receives cavity gradients to predict higher uncertainty for the drifted data.  
Unfortunately, operations staff identified a newly active field emitter affecting the detector at 2L25 between May 9 and May 14.  
Additionally, a cavity was bypassed shortly after taking data on May 9, and re-balancing gradient significantly lowered the radiation for the detector at 2L26.  
It is clear that some level of both concept and data drift are occurring during this work, which complicates the use of UQ.

\section{\label{sec:optimization}Optimization}

The core of our problem is the need to reduce radiation caused by SRF cavity field emission while maintaining the linac energy gain required by the experimental program. 
Linac energy gain can be directly calculated for a given gradient distribution, but predicting the associated radiation levels can only be achieved by use of our surrogate models.
While not all cavities function as inputs to a surrogate model, all do contribute to the linac energy gain and consequently must be included in the optimization problem.

The need to minimize both neutron and gamma radiation while respecting bounds on linac energy gain which defines a multi-objective optimization problem with constraints. 
We also desired flexibility to expand our problem definition to include metrics such as uncertainty quantification or other CEBAF characteristics which may not be readily differentiable. 
This is important for future deployment as radiation production is only one of many problems operations must address when selecting a linac gradient distribution. 
These requirements led us to use an evolutionary algorithm as there are many variants readily available for solving this class of problem. 
We chose the Non-dominated Sorting Genetic Algorithm II (NSGA-II)~\cite{NSGA2} as the main optimization algorithm due to its familiarity within the accelerator physics community at large and Jefferson Lab specifically~\cite{PhysRevSTAB.16.010101,Ge_2023,van_der_geer:ipac15-mopje076}.

Each optimization problem was defined around an initial linac configuration that represented the current operational configuration. 
Bounds on individual cavity gradients were determined as the minimal range that met both real world constraints and a [$-3$, $+0.5$] range around the cavity's initial gradient setting. 
This bounding helps to limit the optimizer from exploring far outside the training distribution, respect real-world cavity limits, and keep solutions close to the initial configuration which addresses many more considerations than simply energy gain and radiation.

Minimizing uncertainty estimates and maximizing energy gain are also included as objectives as they help steer the optimizer in useful directions. 
We prefer regions of data where the observed variation in radiation was smaller, and early studies indicated that prioritizing lower uncertainty values could help prevent the exploitation of model weaknesses.
Linac energy gain is already constrained to an acceptable region for beam transport, but we did not want only solutions clustered at the edge of minimum viability.
Therefore, we include energy gain as both a constraint and a optimization target.

In order to provide a mathematical description of this problem, we first define a few key quantities.
Given a gradient distribution $G=
\begin{smallmatrix}
    [g_1&g_2,&\ldots&g_{200}]
\end{smallmatrix}$, and a vector of cavity lengths $L$, the linac energy gain is calculated as their dot product, \[E(G) = G \cdot L.\]
Different cavity styles have different active lengths, however all are either 0.5 or 0.7 meters. 
As stated earlier, this energy gain must be maintained within an error tolerance $\epsilon$ of the experimentally required value $E^*$.
For our problem, the linac energy gain tolerance is 0.5 \si{}{MeV}.
The predicted neutron radiation and uncertainty for detector $d \in D$ is given as $\hat{y}_t(G,d)$ and $\hat{\sigma}_t(G,d)$, respectively, with radiation type $t\in[n, \gamma]$.
Finally, we define 
\begin{align*}
    R_t(G) &= \sum_{d \in D} \hat{y}_t(G, d) \\
    U_t(G) &= \sum_{d \in D} \hat{\sigma}_t(G, d).
\end{align*}
Given an upper and lower bound for each gradient ($g_i^u$ and $g_i^l$), our optimization problem is given by:
\begin{mini}
  {G}{R_n(G), R_\gamma(G), U_n(G), U_\gamma(G), -E(G)}{}{}
  \addConstraint{|E(G) - E^*|}{\leq \epsilon}
  \addConstraint{ g_i^l \leq g_i \leq g_i^u }{,\quad}{\forall g_i \in G}.
\end{mini}

While the problem is formulated in reducing absolute radiation amounts, the optimization problem should be fairly tolerant to baseline changes in radiation at individual detectors.  
This is because finding the minimum of the function $R_t(G)$ is equivalent to finding the minimum of $R_t(G) + c$ where $c$ is some constant.  
We have noticed a historical pattern of increased or decreased baseline radiation detection over time that does not appear tied to new field emitters.    
While having a surrogate model that can make accurate absolute radiation predictions is ideal, even a model that only captures relative changes in radiation should be sufficient to direct the optimization process to cavity gradients that reduce radiation.

The initial population is generated by a two-phase greedy algorithm. 
The first phase searches for small gradient changes that reduce the total radiation dose rate. 
Each cavity is checked for its radiation reduction potential, and the cavity which corresponds to the largest radiation drop has its gradient reduced for the next iteration of the search. 
The second phase increases cavity gradients in a similar fashion while restoring linac energy gain and increasing the neutron radiation as little as possible. 
This search respects the individual cavity constraints at all steps, but individual iterations may violate the total energy constraint.
The solutions chosen at each iteration of both phases are retained and scaled to meet the energy constraint while respecting individual cavity bounds. 
These solutions form the initial population for NSGA-II. 
This initialization strategy yielded a significant decrease in time to convergence as it both provided a warm start to the multi-objective problem and an initial population that did not violate any constraints.

Convergence criteria is chosen to be a set number of generations due to its simplicity and flexibility. 
Approximate convergence would be achieved on the radiation dose rate objectives after 1,000 - 10,000 generations. 
This corresponds to only a few minutes of compute time on a standard personal computer. 
Minor improvements across all fronts could be achieved if the optimization was run for hours longer. 
However, solutions are often needed in near real-time as the operational environment is dynamic. 
For example, a derated RF cavity immediately requires a new gradient distribution to compensate for its lost energy gain. 
This convergence criteria gave end users control over this time versus optimality trade-off. 
If additional computation time appears needed, the user can simply continue the optimization longer. 
While crude, this approach to convergence provided a simple means to produce radiation reductions in both research and control room situations.

The NSGA-II algorithm produces a family of solutions at various positions along the Pareto front.  
We leverage this in our optimization software to allow operations staff to review the various proposed solutions, and select the one(s) that seem most likely to reduce radiation and respect other accelerator constraints.  
Ultimately, operations staff could simply review the suggested gradient changes and only apply some subset that they believe to be the most workable.  
The flexibility allows for our optimization process to benefit operations in a variety of workflows.

\section{\label{sec:workflow}Results}

In order to provide new gradient settings that reduce FE, we first model the linac radiation response and noise ({\it i.e.} uncertainty) with a neural network based on Simultaneous Quantile Regression.
Then we utilize this surrogate model to optimize radiation with a genetic algorithm subject to a number of constraints including total energy and individual cavity gradient bounds. 
This two stage approach allows analysis of the neural network outputs separately from the optimization process and should produce effective gradient settings given an accurate surrogate model.

We developed python-based software, called FEMinimizer, that provides an interactive GUI for investigation and selection of an optimized gradient distribution. 
Machine learning models are integrated and run using the ONNX runtime framework \cite{onnxruntime}, and optimization algorithms are directly implemented (e.g, greedy) or using the pymoo package (e.g., NSGA-II) \cite{pymoo}.

\begin{figure*}[]
    \centering
    \includegraphics[width=\linewidth]{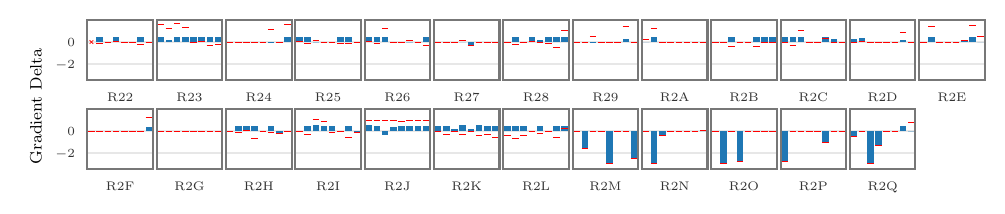}
    \caption{Delta between gradient settings chosen by our NSGA optimizer and the baseline settings (in blue). Since a few settings were modified between the baseline and the beginning of the demonstration, the true demonstration deltas (for \Cref{fig:may19-radiation-responses}) are shown as red lines. R221 was turned off after our baseline was taken, thus this bar is replaced by an ``\textcolor{red}x''. The demonstration delta for R2Q7 was +3.8 (not shown).}
    \label{fig:nsga-solution-gradients}
\end{figure*}

The application allows a user to select a historical point in time and a linac for which to generate an optimization configuration, the cavity gradient offset ranges used in optimization, and the algorithm used to perform the optimization.  
It also features the ability to save, view, or load an optimization configuration.  
Optimizations can be run iteratively for a user-specified number of generations or iterations.  
The set of solutions are displayed as a series of plots showing each pair of objectives.  
For NSGA-II, these are typically called Pareto fronts.  
Individual solutions can be selected based on their location in the  Pareto fronts which loads a more detailed view of that solution.
Once selected, commands to apply that gradient distribution can be generated through the application.  

In order to demonstrate the effectiveness of this integrated approach to FE management we collected data on May 9 and May 14 to support a beam study on May 19. 
As mentioned above, models were trained and validated on May 9 data, tested on May 14 data, and then the trained models were loaded into the FEMinimizer software to generate a population of gradient distributions. 
Below we describe the results of applying one of these solutions to CEBAF.

\subsection{\label{subsec:demo_results}Demonstration Results}

Since the model was still highly correlated to radiation, we conducted a real world test using our NSGA-II optimizer at the next opportunity on May 19.
The optimizer took in the current gradient settings of the south linac, and produced a set of new gradients to optimize for the objectives listed in \Cref{sec:optimization}. 
In order to see how these new settings impacted radiation levels, we applied the largest changes to the linac one-at-a-time, followed by applying all of the smaller changes at once.
The selected changes in gradient settings are shown in \Cref{fig:nsga-solution-gradients}.
Predictably, a few C100 cavity gradients were significantly reduced by the maximum -3 \unit{\mega\volt/\m}, while many C25 cavity gradients were increased to compensate for the lost energy.

Accuracy and correlation results are shown in \Cref{tab:c100-may19-results} with a full plot of the radiation responses in \Cref{fig:may19-radiation-responses}.
Like before, we see larger RMSE errors for most detectors when compared to May 14 (except 2L25 and 2L27) and correlations are still quite good. 
This short test run shows a number of interesting features.
While the model does not track predictions exactly, the major source of error appears to be from an upward shift in radiation.

\begin{table}[]
    \sisetup{round-mode=places, round-precision=2, table-alignment=right, table-format=1.2, table-space-text-pre = $-$}
    \centering \footnotesize
    \caption{Observed and predicted total radiation for our demonstration on May 19. The model predictions are slightly optimistic, but reasonable overall.}
    \begin{tabular*}{0.95\linewidth}{@{\extracolsep{\fill}} lSSSS} \toprule
            & {$y_n$}   & {$\hat{y}_n$} & {$y_\gamma$} & {$\hat{y}_\gamma$} \\ \midrule
    Start   & 18.987425 & 18.625257     & 300.242862   & 240.760369         \\
    End     & 10.477689 & 7.462530      & 176.254873   & 98.303701          \\
    Delta   & 8.509736  & 11.162727     & 123.987989   & 142.456669         \\
    Percent & 44.8177   & 59.9333       & 41.2959      & 59.1695            \\ \bottomrule
    \end{tabular*}     
    \label{tab:may19_total_radiation}
\end{table}

To further verify our results, we show the observed and predicted total neutron and gamma radiation at the start and end of our demonstration, as well as the delta and percentage difference between them in \Cref{tab:may19_total_radiation}. 
At the initial gradient settings ("Start"), the predicted total neutron radiation matches the observed radiation well, unlike total gamma radiation levels which are significantly under-predicted.
The discrepancy in gamma radiation could possibly be explained by \Cref{tab:datasets} which shows total gamma radiation significantly increased between our May 9 training data and the May 19 demonstration, compared to a much smaller increase in total neutron radiation.
It should also be noted that this discrepancy is likely due to changing conditions in the linac, rather than poor model performance since the correlation in the total radiation is 0.991 and 0.988 for neutron and gamma, respectively.
We also see the model is generally optimistic and estimates a 60\% reduction in radiation, exceeding the observed 45\% and 41\% for neutron and gamma, respectively.
Even so, these reductions are significant and represent tangible real-world improvements to radiation management at CEBAF.

\subsection{\label{subsec:follow_results}Fine-Tuning}

\begin{figure*}
    \centering
    \includegraphics[]{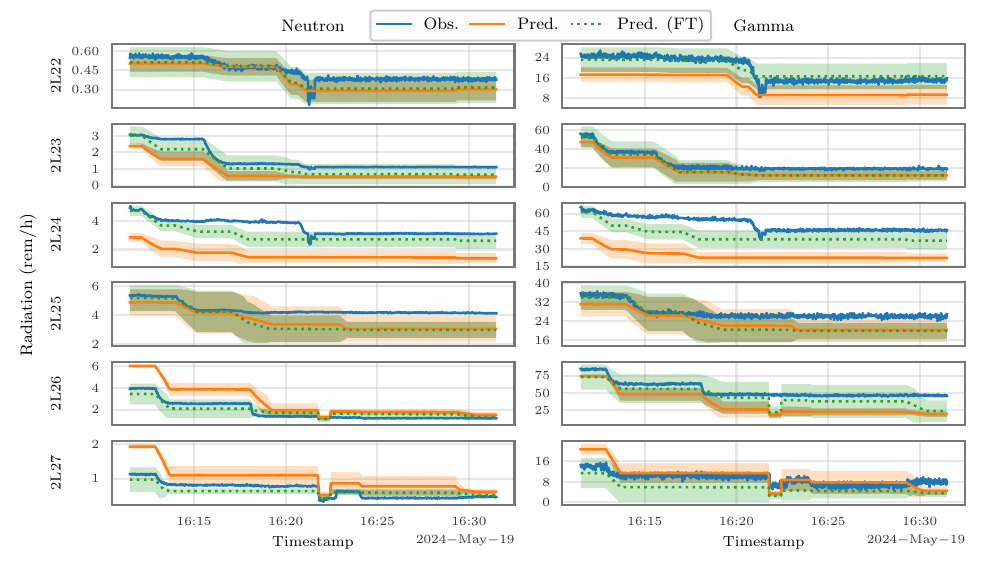}
    \caption{Prediction quality on May 19 demonstration data for our model trained on May 9 and then fine-tuned on the May 19 data. Observations are in blue, predictions are in orange, and the fine-tuned predictions are dotted green. Uncertainties are shown as shaded regions equivalent to $\pm 3 \sigma$. The model is well correlated, and fine-tuning it helps to move predictions towards the observations for most detectors.}
    \label{fig:may19-radiation-responses}
\end{figure*}

Since there was an observable shift in performance between May 9 and our demonstration on May 19, we decided to test the efficacy of fine tuning our model on a small amount of scan data taken on the demonstration day.
To do this, we trained our model for 100 epochs on this smaller, but more recent, dataset. 
As we presented in \Cref{fig:data_drift}, this dataset contains many novel gradient settings, including multiple cavities with changes in baseline gradient settings that were larger than 2 MV/m.
In \Cref{fig:may19-radiation-responses}, we see that the majority of improvement comes from a simple shift in model responses to account for possible detector drifts and gradient changes. 
Since the dataset we tuned the model on was small, and the space of possible gradient settings is very large, it is likely that the fine-tuning dataset does not contain enough information to significantly alter model responses.

The accuracy and correlation results are shown in \Cref{tab:c100-may19-results} and \Cref{fig:may19-radiation-responses} alongside the original model results. 
While the shape of model responses is similar after retraining, we see that it significantly improved 75\% of our RMSE metrics including a reduction of more than 16 and 10 \unit{rem / \hour} for 2L24 $\gamma$ and 2L26 $\gamma$ respectively. 
Only minor degradations in the RMSE performance metric are observed for both 2L25 detectors and the 2L27 $\gamma$ detector. 
Additionally, as seen in \Cref{fig:may19-radiation-responses}, many uncertainty predictions increased, which is likely due to a combination of factors including retained information from the earlier training data, and the broader distribution of residuals from our limited fine tuning.
In general, correlations were also improved, but none of the changes were significant as the model was already highly correlated with the true observations.

\begin{table}[]
    \sisetup{round-mode=places, round-precision=3, table-alignment=right, table-format=1.3, table-space-text-pre = $-$}
    \centering \footnotesize
    \caption{Results of our original model and a fine-tuned model (subscripted $ft$) on the May 19 demonstration data. Highlighted cells indicate better performance. The majority of RMSE results significantly improve when fine-tuning, and the detectors that see worse RMSE scores only decrease by a small amount. }
    \begin{tabular*}{0.95\linewidth}{@{\extracolsep{\fill}} llSSSS} \toprule
    Metric            & Detector & {$n$}           & {$n_{ft}$}      & {$\gamma$}      & {$\gamma_{ft}$}  \\ \midrule
    \multirow{6}{*}{RMSE} & 2L22 & 0.070988        & \optf{0.059038} & 6.482296        & \optf{1.958755}  \\
                          & 2L23 & 0.746222        & \optf{0.416944} & 6.320797        & \optf{5.802085}  \\
                          & 2L24 & 1.977933        & \optf{0.664442} & 26.931548       & \optf{10.539049} \\
                          & 2L25 & \optf{0.840404} & 0.950477        & \optf{4.513756} & 4.926245         \\
                          & 2L26 & 1.048464        & \optf{0.378120} & 20.655354       & \optf{10.581346} \\
                          & 2L27 & 0.372316        & \optf{0.147681} & \optf{2.630558} & 3.704358         \\ \midrule
    \multirow{6}{*}{Corr} & 2L22 & 0.928011        & \optf{0.931916} & \optf{0.969404} & 0.968256         \\
                          & 2L23 & 0.975504        & \optf{0.980638} & 0.986833        & \optf{0.987265}  \\
                          & 2L24 & 0.790083        & \optf{0.798788} & 0.764771        & \optf{0.792054}  \\
                          & 2L25 & 0.883907        & \optf{0.907245} & 0.888767        & \optf{0.901551}  \\
                          & 2L26 & \optf{0.985698} & 0.943632        & \optf{0.977323} & 0.895408         \\
                          & 2L27 & \optf{0.924293} & 0.801647        & \optf{0.842778} & 0.829745         \\ \bottomrule
    \end{tabular*}
    \label{tab:c100-may19-results}
\end{table}

\subsection{\label{subsec:trip_rate}Trip Rate Modeling}

As alluded to in \cref{sec:optimization}, operational gradient distributions are optimized against many additional considerations. 
A primary concern of the current gradient distribution methods is minimizing the number of RF cavity arc trips that occur within the linac ~\cite{benesch2015festudy,PhysRevAccelBeams.19.124801}. 
An individual cavity's arc trip rate is well modeled using a log-linear statistical relationship with its gradient, and these model parameters are stored in a historical database similar to other CEBAF parameters. 
The estimated trip rate across the linac is simply the sum of the trip rates of the modeled cavities.  
A critical follow-up question to the May 19 demonstration is how redistributing gradient away from active field emitters impacts the overall trip rate.

In order to investigate this topic, we modify the optimization problem from \cref{sec:optimization} by adding the minimization of the linac trip rate as an objective, including a constraint bounding the maximum linac trip rate at ten trips/hour, and relaxing the individual cavity gradient change constraints from +0.5 MV/m to +1.0 MV/m. 
We found that limiting gradient increases to only 0.5 MV/m had a significant impact on trip rates as the optimizer was forced to increase gradients on less stable cavities. 
This is consistent with the fact that trip rates are modeled using an exponential response to cavity gradients and some cavities are much more prone to trips than others.

\begin{table}
    \sisetup{round-mode=places, round-precision=2, table-alignment=right, table-format=3.2}
    \centering \footnotesize
    \caption{Comparison of model estimated radiation and trip rates between the demonstration starting point (Baseline), the optimized gradients from the demonstration (Optimized), and gradients re-optimized for  both trip rate and radiation levels (Re-Optimized).}
    \begin{tabular*}{0.95\linewidth}{@{\extracolsep{\fill}} lSSS} \toprule
                       & {Baseline} & {Optimized} & {Re-Optimized} \\ \midrule
    Trip Rate (trip/h) & 4.27       & 6.92        & 4.27           \\
    Neutron (rem/h)    & 18.63      & 7.46        & 11.73          \\
    Gamma (rem/h)      & 240.76     & 98.30       & 159.12         \\ \bottomrule
    \end{tabular*}
    \label{tab:trip_rates}

\end{table}

Using this new optimization problem, we generated a population of gradient distributions and highlight a specific re-optimized gradient distribution that yields a trip rate similar to the baseline (\cref{tab:trip_rates}). 
This new solution suggests a neutron and gamma radiation reduction of 37\% and 34\%, respectively. 
Unfortunately it is impossible to know what the real radiation would be for this new solution. 
Given the observed tendency of our models to overestimate radiation reduction for large gradient changes we suspect that these estimates may also be optimistic.  
However, that suspicion is tempered by this new solution's more conservative gradient reductions.
That makes it reasonable to believe that this solution would produce substantial radiation reductions even if not at the previously observed levels of 45\% and 41\% for neutron and gamma respectively. 
It is important to note that even if this estimate is true, the inclusion of additional operational considerations could increase radiation and/or trips by further constraining the problem.

\section{Discussion and Conclusion}

Our final demonstration on May 19 gives a proof of concept about how this approach could benefit operations through reduced radiation. 
Operations staff had been mitigating radiation throughout the experimental run prior to our demonstration, therefore the 40-45\% reduction in radiation represents a real, tangible decrease in radiation over what would have been seen in operations. 
Field emission has been a constant problem, and our approach represents a viable method to significantly reduce its observed effects through passive mitigation.

As we note throughout this work, CEBAF operations contends with many considerations beyond field emission when setting cavity gradients. 
Operations must also address cavity trip rates, electrical power limits, RF klystron power limits, energy lock gradient reserves, amongst others concerns. 
Integrating field emission mitigation into linac gradient distribution requires a considerable effort and system expert discretion. 
Trip rates represent one of the most pressing concerns when optimizing gradient distributions, and our brief investigation suggests that significant radiation reduction can be achieved even when accounting for cavity trips.

Our model architecture provides for reasonable surrogate models. 
It avoids many extraneous correlations present in our relatively small training data sets and can be readily fine-tuned on new data. 
We expect that further refining data collection practices would improve the already sufficient model performance.
While our model provides uncertainty quantification, this aspect suffers from drift in the configuration of CEBAF and the presence of at least one new active field emitter.
As such, it appears that detecting OOD conditions is an open question in this challenging scenario.
However, our model appears to do a reasonable job of representing the uncertainty introduced by random fluctuations in detector noise.
Future improvements in OOD detection could help dissuade optimization processes from exploiting model weaknesses due to the limited training data and the large input space.
Another avenue for improvement would be to model the entire linac with a single model. 
Most of our models were simple by design to avoid spurious correlations that often appeared in our data.
Future work could investigate other techniques such as self-attention to better identify which cavities are field emitters, physics-informed models to ensure model outputs better represent the real world, and methods for uncertainty quantification that improve out-of-distribution identification.

For the model to remain reflective of radiation response through the span of an experimental run, we would likely need to institute continual learning techniques.
Here, we take the first successful step in this direction, by showing the ability for fine tuning on a small training set to reduce prediction errors. 
This data was collected in 30 minutes and could reasonably be performed by operations semi-routinely. 
An initial step in continual learning could involve retraining only the final biases for models with good correlation but poor errors due to shifts.

Another avenue for maintaining model accuracy over time is leveraging passive data from CEBAF.  RF cavities routinely trip and recover giving potentially useful information on how radiation is impacted by changing gradient.  Additionally, cavity gradients are routinely redistributed throughout the linacs in order to maintain a constant linac energy gain.  Using these passive data sources may reduce our reliance on invasive data collection that interferes with experimental runs.  We briefly explored this topic in \cite{ahammed:ipac2024-mopc44}, however that work focused on only a single section of a linac and was hampered by data limitations.  This work did suggest that RF trip data could be useful in improving model performance in certain cases, making further study warranted. 

Many tools were created to enable this work, and most represent a solid start on production software needed to implement this in routine operations. 
While data collection still requires some amount of manual intervention, we have developed a software framework for controlling RF throughout a linac and implementing a variety of sampling strategies. 
The optimization software already provides a polished user interface and is integrated with CEBAF systems for accurately reflecting CEBAF configuration at arbitrary points in time. 
This significantly lowers the barrier to entry for production use of the techniques showcased in this work.

\section{Acknowledgements}

We would like to acknowledge Jay Benesch, Pavel Degtiarenko, and Michael McCaughan for their many informative conversations, and CEBAF operations and engineering staff more broadly for their support throughout this endeavor.
This work is supported by the U.S. Department of Energy, Office of Science, Office of Nuclear Physics under Contract No. DE-AC05-06OR23177.
\vspace{4.0em}
\bibliography{fe_paper}

\appendix
\vspace{-0.2em}
\section{Additional Results}
\vspace{-0.2em}
\label{appendix:2L04/17_Results}

Here, we present additional results for the non-C100 detectors. 
Compared to the C100 detectors, these detectors exhibit consistently low error but poor correlations.
Since these detectors have almost no true signal and are dominated by noise, any model with good correlation results would be over-fitting to the true data.
The only exception are the results for 2L08 radiation which sometimes contains a change in radiation signal that can be modeled. 

\Cref{tab:full-may14-results} presents validation results of our model on the May 14 dataset.
\Cref{tab:full-may19-results} shows the same metrics (RMSE and Pearson's correlation) for the May 19 demonstration.
\Cref{tab:full-fine-tuning-on-may19} shows the results of the fine-tuned model on the May 19 demonstration dataset.

\begin{table}[b]
    \sisetup{round-mode=places, round-precision=3, table-alignment=center, table-format=1.3, table-space-text-pre = $-$}
    \centering \footnotesize
    \caption{RMSE and correlation on the May 14 dataset for our model trained on the full May 9 data.}
    \begin{tabular*}{0.95\linewidth}{@{\extracolsep{\fill}} lSSSS} \toprule
    & \multicolumn{2}{c}{RMSE} & \multicolumn{2}{c}{Corr} \\
    Detector & {n}      & {$\gamma$} & {n}      & {$\gamma$} \\ \midrule
    2L04     & 0.008220 & 0.576253   & 0.098316 & 0.364538   \\
    2L07     & 0.007355 & 0.706230   & 0.027537 & 0.371420   \\
    2L08     & 0.016516 & 1.142217   & 0.721004 & 0.873572   \\
    2L10     & 0.013835 & 0.510995   & 0.026705 & -0.022932  \\
    2L11     & 0.013526 & 1.997674   & 0.014103 & -0.057093  \\
    2L16     & 0.017319 & 0.940347   & 0.023815 & 0.263034   \\
    2L17     & 0.017082 & 2.239649   & 0.186051 & 0.081372   \\ \bottomrule
    \end{tabular*}
    \label{tab:full-may14-results}
\end{table}

\begin{table}[]
    \sisetup{round-mode=places, round-precision=3, table-alignment=center, table-format=1.3, table-space-text-pre = $-$}
    \centering \footnotesize
    \caption{RMSE and correlation on the May 19 dataset for our model trained on the full May 9 data.}
    \begin{tabular*}{0.95\linewidth}{@{\extracolsep{\fill}} lSSSS} \toprule
    & \multicolumn{2}{c}{RMSE} & \multicolumn{2}{c}{Corr} \\
    Detector & {n}      & {$\gamma$} & {n}       & {$\gamma$} \\ \midrule
    2L04     & 0.008178 & 0.512666   & 0.003424  & 0.069030   \\
    2L07     & 0.007820 & 0.671279   & -0.097836 & -0.352951  \\
    2L08     & 0.011419 & 2.906832   & 0.038802  & 0.793208   \\
    2L10     & 0.013824 & 0.491809   & 0.052618  & -0.042063  \\
    2L11     & 0.013619 & 1.967242   & -0.070917 & 0.057915   \\
    2L16     & 0.016468 & 0.861172   & -0.016672 & 0.006842   \\
    2L17     & 0.016343 & 2.255221   & -0.033751 & 0.009505   \\ \bottomrule
    \end{tabular*}
    \label{tab:full-may19-results}
\end{table}

\begin{table}[H]
    \sisetup{round-mode=places, round-precision=3, table-alignment=center, table-format=1.3, table-space-text-pre = $-$}
    \centering \footnotesize
    \caption{RMSE and correlation results on the May 19 demonstration set using a fine-tuned May 9 model. The fine-tuning utilized data collected on May 19, prior to the demonstration. }
    \begin{tabular*}{0.95\linewidth}{@{\extracolsep{\fill}} lSSSS} \toprule
    & \multicolumn{2}{c}{RMSE} & \multicolumn{2}{c}{Corr} \\
    Detector & { n}     & { $\gamma$} & { n}      & { $\gamma$} \\ \midrule
    2L04     & 0.008087 & 0.514242    & 0.003365  & 0.069017    \\
    2L07     & 0.007741 & 0.588361    & -0.093417 & -0.352444   \\
    2L08     & 0.012154 & 1.090945    & 0.039493  & 0.798126    \\
    2L10     & 0.013918 & 0.493618    & 0.051126  & -0.042603   \\
    2L11     & 0.013621 & 1.962115    & -0.070167 & 0.055888    \\
    2L16     & 0.016308 & 0.857180    & -0.016451 & 0.006938    \\
    2L17     & 0.016267 & 2.255190    & 0.032110  & -0.012837   \\
    \bottomrule
    \end{tabular*}
    \label{tab:full-fine-tuning-on-may19}
\end{table}

\end{document}